# Higher-order geometric phase for qubits in a bichromatic field

A P Saiko[1], R Fedaruk[2] and A Kolasa[2]

[1]Scientific-Practical Materials Research Centre NAS of Belarus, Minsk, Belarus

[2]Institute of Physics, University of Szczecin, 70-451, Szczecin, Poland

E-mail: saiko@ifttp.bas-net.by; fedaruk@wmf.univ.szczecin.pl

The geometric phase in the dynamics of a spin qubit driven by transverse microwave (MW) and longitudinal radiofrequency (RF) fields is studied. The phase acquired by the qubit during the full period of the "slow" RF field manifests in the shift of Rabi frequency $\omega_1$ of a spin qubit in the MW field. We find out that, for a linearly polarized RF field, this shift does not vanish at the second and higher even orders in the adiabaticity parameter $\omega_{rf}/\omega_1$, where $\omega_{rf}$ is the RF frequency. As a result, the adiabatic (Berry) phases for the rotating and counter-rotating RF components compensate each other, and only the higher-order geometric phase is observed. We experimentally identify that phase in the frequency shift of the Rabi oscillations detected by a time-resolved electron paramagnetic resonance.

## 1. Introduction

Berry has shown [1] that during adiabatic and cyclic evolution of Hamiltonian's parameters, a non-degenerate eigenstate of a quantum system acquires a geometric phase in addition to its dynamic counterpart. The concept of the geometric phase was extended to non-cyclic and nonadiabatic (*e.g.*, the Aharonov-Anandan phase [2]) evolutions, including degenerate states [2] and open systems with dissipation [3]. Geometric phases manifest in a wide variety of phenomena in quantum physics, optics, and solid state physics (see, *e.g.*, [2,4]). For example, the realizations of exotic (quantum [5], anomalous [6], optical [7], and phononic [8]) Hall effects as well as new propositions of logic gates for quantum computation [9] are closely related to the geometric phase. Various types of basic elements (qubits) for quantum computation have been presented in recent reviews [10, 11].

Energy level shifts of a quantum system due to off-resonant periodic perturbations, such as Bloch-Siegert, dynamic (ac) Zeeman and Stark shifts, are also directly related to the additional phase in the evolution of its stationary states. Indeed, slow periodic adiabatic perturbations with periods $\tau$ much longer than the characteristic time $\tau_s$ of the quantum system's evolution, give rise to energy shifts including that proportional to the first order in the adiabaticity parameter $\tau_s/\tau$ (see, *e.g.*, [12]).

This ensured that the quantum system accumulates the geometric phase [12]. In some cases [13], the energy level shift does not appear in the first order in $\tau_s/\tau$. As a result, the geometric phase can be acquired only at higher orders in $\tau_s/\tau$, *i.e.* when the adiabaticity condition is violated.

The total phase $\phi$ acquired by the wavefunction of a quantum system in the adiabatically changing field can be expanded in a power series in a small parameter $\varepsilon = \tau_s/\tau$:

$$\frac{d\phi}{dt} = \alpha_0 \varepsilon^0 + \alpha_1 \varepsilon^1 + \alpha_2 \varepsilon^2 + ... \qquad (1)$$

The dynamical phase of the wavefunction is given by the integral of the zeroth-order term over the time interval (0, $\tau$). The integral of the first-order term depends on the geometry of the closed path in the space of slowly changing parameters and corresponds to the geometric (Berry) phase. Integrals of higher-order terms yield corrections to the Berry phase and become zero in the exact adiabatic limit $\varepsilon \to 0$. However, for real physical processes, adiabatic conditions are often not exactly fulfilled, i.e. the adiabaticity parameter has a small but finite magnitude. In these cases, higher-order corrections in $\varepsilon$ are needed (see [14] and the references therein, and also [11]). Note that, in order to weaken the fulfillment of the exact adiabatic condition, Berry has also performed calculations taking into account the adiabatic correction terms (see p.494 in [2]). Examples of the higher-order corrections to the geometric phase for simple systems can be found in literature, *e.g.* in [15, 16]. The higher-order corrections to the geometric phase [15] have been used to explain phase fluctuations observed in the superconducting circuit system [17]. Even for simple quantum systems such as spin qubits in a slowly changing electromagnetic field, the Berry phase, in its standard definition, exists only when the field is circularly polarized [18]. For the linearly polarized field the first-order term in $\varepsilon$ in equation (1) that describes the standard Berry phase becomes zero, but the even higher-order terms in the adiabaticity parameter differ from zero. Consequently, the geometric phase observed in that case can originate only from the higher-order corrections. We call the higher-order geometric phase the part of the geometric phase given by the higher-order corrections when the Berry phase is zero. The higher-order geometric phase cannot be treated as the Aharonov-Anandan phase, because for Hermitian driving Hamiltonians the last one must reduce to the Barry phase in the adiabatic limit.

On the other hand, beyond the adiabatic limit, the periodically driven two-level system can undergo non-adiabatic multiple Landau-Zener transitions resulting in the so-called Landau-Zener-Stückelberg (LZS) interference [19]. Recently, the link between the LZS interference and geometric phases, which opens new possibilities for geometric control of quantum systems, has been considered [20]. In Ref [21], it has been demonstrated that a Berry phase evolving linearly in time induces a frequency shift of the resonance transition between two eigenstates.

In the present paper, we study the higher-order geometric phase for a spin qubit driven by the "fast" microwave (MW) and "slow" radiofrequency (RF) fields. The bichromatic excitation with the linearly polarized RF field allows us to eliminate the Berry phase of the qubit and to observe only the higher-order geometric phase. We construct the effective Hamiltonian of the quantum system with help of the Krylov-Bogoliubov-Mitropolsky non-secular perturbation theory and then solve the Liouville equation for the density matrix including dissipative processes. We predict the appearance of the higher-order geometric phase and experimentally identify that phase in the frequency shift of the Rabi oscillations using electron paramagnetic resonance (EPR) for the $E'_1$ centers in quartz.

## 2. Theory

We consider a spin qubit in a MW field oriented along the $x$ axis of the laboratory frame, in presence of the RF and static magnetic fields, both directed along the $z$ axis. The Hamiltonian of the qubit in these fields can be written as:

$$H = H_0 + H_\perp(t) + H_\parallel(t), \qquad (2)$$

where $H_0 = \omega_0 s^z$ is the Hamiltonian of the Zeeman energy of the qubit in the static magnetic field $B_0$, $\omega_0 = \gamma B_0$, $\gamma$ is the gyromagnetic ratio; whereas $H_\perp(t) = \omega_1(s^+ + s^-)\cos\omega_{mw}t$ and $H_\parallel(t) = 2\omega_2 s^z \cos\omega_{rf} t$ are the Hamiltonians of the qubit interaction with linearly polarized MW and RF fields, respectively. Here $B_1$ and $B_2$, $\omega_{mw}$ and $\omega_{rf}$ are the respective amplitudes and frequencies of the MW and RF fields. Moreover, the Rabi frequencies $\omega_1 = \gamma B_1$ and $\omega_2 = \gamma B_2$ denote the respective interaction constants for the MW and RF fields, whereas $s^{\pm,z}$ are components of the spin operator, describing the state of the qubit and satisfying the commutation relations: $[s^+, s^-] = 2s^z$, $[s^z, s^\pm] = \pm s^\pm$. In addition, we consider the "dressed" states of the qubit at the precise resonance between the MW field and the qubit ($\omega_{mw} = \omega_0$), also assuming that $\omega_0 \gg \omega_1 \gg \omega_{rf}$. Since in our experiment $\omega_1/\omega_{mw} \approx 10^{-4}$, the counter-rotating component of the MW field is neglected and the rotating-wave approximation is used for the interaction between the qubit and the MW field. For the RF field, the rotating and counter-rotating components should be taken into account because $\omega_2$ и $\omega_{rf}$ can be comparable.

Dynamics of the system with the Hamiltonian (2) is described by the Liouville equation for the density matrix $\rho$:

$$i\hbar \frac{\partial \rho}{\partial t} = [H, \rho] + i\Lambda\rho \qquad (3)$$

(further we assume $\hbar = 1$). The superoperator $\Lambda$ describing decay processes is defined by its action upon ρ as:

$$\Lambda\rho = (\gamma_{21}/2)(2s^-\rho s^+ - s^+s^-\rho - \rho s^+s^-) + (\gamma_{12}/2)(2s^+\rho s^- - s^-s^+\rho - \rho s^-s^+) + (\eta/2)(2s^z\rho s^z - \rho/2),$$

where $\gamma_{21}$ and $\gamma_{12}$ are the rates of the transitions from the excited state 2 of the qubit to its ground state 1 and vice versa, and $\eta$ is the dephasing rate. After two canonical transformations $\rho_2 = u_2^+ u_1^+ \rho u_1 u_2$, where $u_1 = \exp(-i\omega_{mw}ts^z)$, $u_2 = \exp(-i\pi s^y/2)$, equation (3) is transformed into $i\partial\rho_2/\partial t = [H_2, \rho_2] + i\Lambda'\rho_2$ if the conditions $\gamma_{21}, \gamma_{12}, \eta \ll \omega_1$ are fulfilled. Here

$$H_2 = \omega_1 s^z - \frac{\omega_2}{2}(s^+ e^{-i\omega_{rf}t} + h.c.) - \frac{\omega_2}{2}(s^+ e^{i\omega_{rf}t} + h.c.), \quad (4)$$

where the second and third terms in the Hamiltonian (4) correspond to the rotating wave (RW) and counter-rotating wave (CRW) interactions between the RF field and the spin qubit, respectively, $\Lambda' \rho_2$ has the same operator structure as $\Lambda \rho$ and differs only in the relaxation parameters: $\gamma_{21} \to \Gamma_\downarrow = (\gamma_{21} + \gamma_{12} + \eta)/4$, $\gamma_{12} \to \Gamma_\uparrow = \Gamma_\downarrow$, $\eta \to \Gamma_\varphi = \gamma_{21} + \gamma_{12}$. The operator $u_1$ transforms the density matrix in the frame that rotates with the frequency $\omega_{mw}$ around the $z$ axis of the laboratory frame, and $u_2$ tilts that frame in the spin space by the angle $\pi/2$ around the $y$ axis. Note that, when the condition $\omega_{rf} \ll \omega_1$ is fulfilled, the probabilities of the RW and CRW processes are close to each other in magnitude.

Using the transformation operator $u_3 = \exp(-i\omega_1 t s^z)$, the Liouville equation can be transformed in the doubly rotating frame which rotates with the frequency $\omega_1$ around the $z$ axis of the tilted frame. In this frame in the Hamiltonian (4) the oscillations with the Rabi frequency $\omega_1$ occur. Since $\omega_1 \gg \omega_{rf}$, the ratio $\omega_{rf}/\omega_1$ can be used as the adiabaticity parameter. Hence, the Hamiltonian (4) contains rapidly $\exp(\pm i\omega_1 t)$ and slowly $\exp(\pm i\omega_{rf} t)$ oscillating functions. The rapidly oscillating terms in the transformed Liouville equation can be eliminated by using the Krylov–Bogoliubov–Mitropolsky non-secular perturbation theory [22, 23, 24]. In the second order in a small parameter $\omega_2/\omega_1$, we get,

$$i\partial\langle\rho_3\rangle/\partial t = \left[H_{eff}, \langle\rho_3\rangle\right] + i\langle\Lambda'\rangle\langle\rho_3\rangle,$$

$$H_{eff} = \frac{i}{2}\left\langle\left[\int^t d\tau\left(H_3(\tau) - \langle H_3(\tau)\rangle\right), H_3(t)\right]\right\rangle = \delta\Omega^{(2)}(t)s^z, \quad (5)$$

where $\rho_3 = u_3^+ \rho_2 u_3$, the symbol $\langle...\rangle$ denotes time averaging over rapid oscillations of the type $\exp(\pm i\omega_1 t)$ given by $\langle O(t)\rangle = \frac{\omega_1}{2\pi}\int_0^{2\pi/\omega_1} O(t)dt$ and the upper limit $t$ of the indefinite integral indicates the variable on which the result of the integration depends, and square brackets denote the commutation operation. The averaging does not affect the form of the superoperator, $\langle\Lambda'\rangle = \Lambda'$. The frequency shift in the Rabi oscillations is given by

$$\delta\Omega^{(2)}(t) = \frac{1}{2}\omega_2^2\left(\frac{1}{\omega_1 - \omega_{rf}} + \frac{1}{\omega_1 + \omega_{rf}}\right)(1 + \cos 2\omega_{rf}t). \quad (6)$$

Since, in the doubly rotating frame, the phase $\phi^{(2)}$ acquired by the qubits equals $\phi^{(2)}(t) = \int \delta\Omega^{(2)} dt$, using equation (1) and expanding $\delta\Omega^{(2)}$ (6) in a power series of $\omega_{rf}/\omega_1$ yields $\delta\Omega^{(2)} = \delta\Omega_d^{(2)} + \delta\Omega_g^{(2)}$, where $\delta\Omega_d^{(2)} = (\omega_2^2/\omega_1)(1 + \cos 2\omega_{rf}t)$ is the dynamical Zeeman shift, and $\delta \Omega_g^{(2)} = (\omega_2^2 \omega_{rf}^2 / \omega_1^3)(1+\cos 2\omega_{rf} t)$ is the shift corresponding to the geometric phase. The latter shift is quadratically dependent on the adiabaticity parameter $\omega_{rf}/\omega_1$ (even higher-order terms are neglected).

After a full cycle $\tau = 2\pi/\omega_{rf}$ of the "slow" RF field's evolution the qubit accumulates the dynamic phase $\phi_d^{(2)} = 2\pi\omega_2^2/\omega_1\omega_{rf}$ and the geometric phase $\phi_g^{(2)} = 2\pi\omega_2^2\omega_{rf}/\omega_1^3$. The geometric phase is the first order term in the adiabaticity parameter and tends to zero at the adiabatic limit of $\omega_{rf}/\omega_1 \to 0$. Consequently, we get $\delta\Omega_g^{(2)} = (\phi_g^{(2)}\omega_{rf}/2\pi)(1+\cos 2\omega_{rf}t)$.

The terms $(\omega_1 - \omega_{rf})^{-1}$ and $(\omega_1 + \omega_{rf})^{-1}$ in equation (6) are due to the RW and CRW components of the RF field in the Hamiltonian $H_3$ (4), respectively. The "interference" between these components is described by the term containing $\cos 2\omega_{rf}t$ in the shift $\delta\Omega^{(2)}$. The "interference" contribution is eliminated by averaging over the period $2\pi/\omega_{rf}$, and we obtain $\overline{\delta\Omega^{(2)}} = \omega_2^2/\omega_1 + \phi_g^{(2)}\omega_{rf}/2\pi$. The higher-order (at least the fourth-order) corrections in $\omega_2/\omega_1$ allow us to improve the theoretical description. The application of the non-secular perturbation theory [22] gives the fourth-order correction to the shift of the Rabi frequency in the form: $\overline{\delta\Omega^{(4)}} = -\omega_2^4/4\omega_1^3 - 3\omega_2^4\omega_{rf}^2/2\omega_1^5$. When both the second-order and the fourth-order corrections are considered, the shift becomes

$$\overline{\delta\Omega} = \overline{\delta\Omega^{(2)}} + \overline{\delta\Omega^{(4)}} = \frac{\omega_2^2}{\omega_1}\left(1 - \frac{\omega_2^2}{4\omega_1^2}\right) + \frac{\omega_2^2 \omega_{rf}^2}{\omega_1^3}\left(1 - \frac{3\omega_2^2}{2\omega_1^2}\right), \quad (7)$$

and the geometric phase is

$$\phi_g = \phi_g^{(2)} + \phi_g^{(4)} = \frac{2\pi\omega_2^2}{\omega_1^2}\left(1 - \frac{3\omega_2^2}{2\omega_1^2}\right)\frac{\omega_{rf}}{\omega_1}. \qquad (8)$$

The results of the non-secular perturbation theory lead to the important conclusion. Because the "interference" of the RW and CRW terms in the frequency shift and in the geometric phase is averaged to zero over the period $2\pi/\omega_{rf}$, the contributions of these terms in the Hamiltonian (4) can be calculated independently in all orders in $\omega_2/\omega_1$. Therefore, it is possible to consider the RW ($H_2^{RW}$) and CRW ($H_2^{CRW}$) parts in the Hamiltonian (4) separately. Each of these parts is exactly diagonalized: $H_2^{RW} = f(\omega_{rf})s^z$, $H_2^{CRW} = f(-\omega_{rf})s^z$, where $f(\omega_{rf}) = \left[(\omega_1 - \omega_{rf})^2 + \omega_2^2\right]^{1/2}$. Taking into account in the same way the contributions of the RW and CRW components of the RF field, we calculate the shift of the Rabi frequency up to the second order in $\omega_{rf}/\omega_1$. In the arbitrary order in $\omega_2/\omega_1$ we have:

$$\overline{\delta\Omega} = 2(f(0) - \omega_1) + \omega_2^2 f^{-3}(0)\omega_{rf}^2. \qquad (9)$$

equation (9) is "exact" in comparison with equation (7) because it is correct for all ratios of $\omega_2/\omega_1$. The geometric phase corresponding to the frequency shift (9) is

$$\phi_g = 2\pi\omega_2^2 f^{-3}(0)\omega_{rf}. \qquad (10)$$

Hence, equations (9) and (10) are generalizations of equations (7) and (8) calculated by using the perturbation theory for the case of the arbitrary ratio of $\omega_2/\omega_1$.

Figure 1 shows the qubit dynamics on the Bloch sphere in the frame rotating with the frequency $\omega_{mw}$ and tilted by the angle $\pi/2$ around the $y$ axis in the spin space. The vectors $\boldsymbol{\omega}_{rf}$ and $-\boldsymbol{\omega}_{rf}$ represent the rotating and counter-rotating RF field. At the exact resonance ($\omega_{mw} = \omega_0$) in the rotating wave approximation for the MW field, the vector of the qubit magnetization $\boldsymbol{\mu}$ can precess about the vectors $\boldsymbol{\omega}_+$ and $\boldsymbol{\omega}_-$ in the plane perpendicular to that vectors. The vectors $\boldsymbol{\omega}_+$ and $\boldsymbol{\omega}_-$ rotate in the opposite directions around $\boldsymbol{\omega}_1$ at the respective angles $\theta - \theta'$ and $\theta + \theta''$ relative to $\boldsymbol{\omega}_1$. Here $\theta = \arccos[\omega_1/f(0)]$. Assuming that $\omega_{rf}/\omega_1 \ll 1$ and $\omega_2/\omega_1 \ll 1$, $\theta' \approx \theta''$.

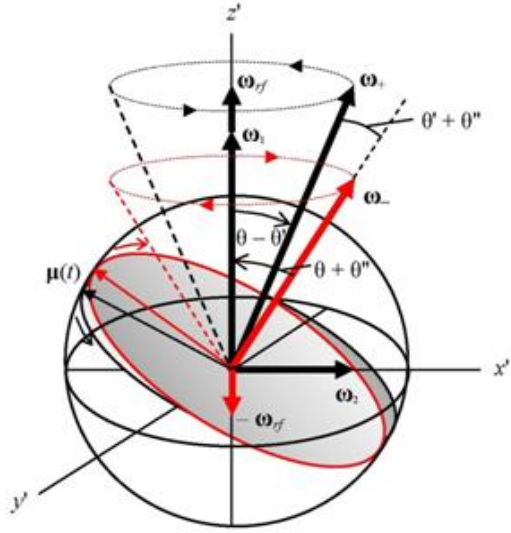

Figure 1. The dynamics of the magnetization vector $\boldsymbol{\mu}$ of the qubit on the Bloch sphere in the frame rotating with the frequency $\omega_{mw}$ and tilted by the angle $\pi/2$ around the $y$ axis in the spin space. The vectors $\boldsymbol{\omega}_{rf}$ and $-\boldsymbol{\omega}_{rf}$ represent the rotating and counter-rotating components of the RF field. The vectors of the effective frequencies $\boldsymbol{\omega}_+$ and $\boldsymbol{\omega}_-$ rotate in the opposite directions around $\boldsymbol{\omega}_1$ at the respective angles $\theta - \theta'$ and $\theta + \theta''$ relative to $\boldsymbol{\omega}_1$. The magnetization vector can precess about the vectors $\boldsymbol{\omega}_+$ and $\boldsymbol{\omega}_-$ in the plane perpendicular to those vectors.

The effective frequencies of the qubit in the rotating and counter-rotating RF field are given by $\omega_- = f(\omega_{rf})$ and $\omega_+ = f(-\omega_{rf})$, respectively. The geometric phase acquired by the qubit state after a full cycle of the evolution, which is described by the Hamiltonian $H_2^{RW}$ ($H_2^{CRW}$), equals $+\Theta_{C-}/2$ ($-\Theta_{C+}/2$), where $\Theta_{C-}$ ($\Theta_{C+}$) is the solid angle enclosed by the loop traversed by $\boldsymbol{\omega}_-$ ($\boldsymbol{\omega}_+$). The solid angles are given by $\Theta_{C-} = 2\pi[1-\cos(\theta+\theta')]$ and $\Theta_{C+} = 2\pi[1-\cos(\theta-\theta')]$, where $\cos(\theta+\theta') = (\omega_1 - \omega_{rf})f^{-1}(\omega_{rf})$ and $\cos(\theta-\theta') = (\omega_1 + \omega_{rf})f^{-1}(-\omega_{rf})$. Expanding $\Theta_{C+}$ and $\Theta_{C-}$ in the power series of $\omega_{rf}/\omega_1$ up to the first order term, we take into account the higher-order effect and obtain $\Theta_{C-} - \Theta_{C+} = 4\pi\omega_2^2 \omega_{rf} f^{-3}(0) \equiv \Theta'_C$. The difference $\Theta'_C$ has the finite value at the first order in the adiabaticity parameter $\omega_{rf}/\omega_1$ and becomes zero at ideal adiabaticity. In addition to the MW and RF amplitudes, the solid angle $\Theta'_C$ depends on the RF frequency. According to equation (10) the higher-order geometric phase is $\varphi_g = \Theta'_C/2$. Hence, the higher-order geometric phase is determined by the half of the solid angle obtained by the difference between the solid angles corresponding to the the RW and CRW components of the RF field.

The density matrix in the laboratory frame is given by $\rho(t) = u_1 u_2 u_3 \exp\left(-i\int \delta\Omega s^z dt\right)\left(\exp(\Lambda_3 t)(u_2^+ \rho(0) u_2)\right)\exp\left(i\int \delta\Omega s^z dt\right) u_3^+ u_2^+ u_1^+$, where $\rho(0) = 1/2 - s^z$ provided that the qubit is found in the ground state at the initial moment. By using $\rho(t)$, we obtain the absorption signal $V$ of the qubit in the rotating frame:

$$V(t) = tr\left\{\rho(t) s^y\right\} = \frac{1}{2i}(\langle 1|\rho(t)|2\rangle - \langle 2|\rho(t)|1\rangle) = \frac{1}{2}\exp(-\Gamma_\perp t)\sin\left(\omega_1 t + \int \delta\Omega dt\right), \quad (11)$$

where $|1\rangle$ and $|2\rangle$ are the ket-vectors of the ground and excited states of the qubit, $\Gamma_\perp = (\Gamma_\downarrow + \Gamma_\uparrow + \Gamma_\varphi)/2$. The renormalized relaxation rates $\Gamma_\perp$ can be expressed in terms of the longitudinal $T_1$ and transverse $T_2$ relaxation times in the laboratory frame: $\Gamma_\perp = 1/2T_2 + 1/2T_1$.

If the system of qubits has a distribution of its own frequencies $\omega_0^i$ (where the index $i$ numbers the qubit), the absorption signal of the system $V_\Sigma$ can be found by summing up signals of every qubit with an appropriate weight function [25]. Assuming that $\Gamma_\perp$ and $\delta\Omega$ are the same for all qubits and that the width of the weight function is much larger than $\omega_1$, we obtain:

$$V_\Sigma \sim \omega_1 \exp(-\Gamma_\perp t)\left[J_0(\omega_1 t)\cos\left(\int \delta\Omega dt\right) - N_0(\omega_1 t)\sin\left(\int \delta\Omega dt\right)\right], \quad (12)$$

where $J_0(x)$ and $N_0(x)$ are the zero order Bessel function of the first and second kind, respectively.

## 3. Experimental results and discussion

Rabi oscillations were formed by the so-called Zeeman pulse technique [26]. The continuous MW and RF fields were used. The resonant interaction with the spin system was abruptly established with a pulse of a longitudinal magnetic field. Pulses of a magnetic field with amplitude $\Delta B = B - B_0 =$ 0.12 mT and duration of 10 μs were used. The experiments were performed at room temperature using an X-band home-made EPR spectrometer. Multichannel digital summation was used to improve the signal-to-noise ratio. Due to the long relaxation times and the small EPR line width, the $E'_1$ centers ($S$ = 1/2) in neutron-irradiated quartz were used in our experiments. The EPR spectrum of the $E'_1$ centers consists of an isolated line with the width of $\Delta B_{pp}$ = 16 μT at the magnetic field direction along the crystal optical axis. The values of $\omega_1$ and $\omega_2$ were directly measured with the precision of about 1 kHz at $\omega_{mw} = \omega_0$ using the frequencies of the Rabi oscillations in the MW field ($\omega_2 = 0$) and in the bichromatic field at $\omega_{rf} = \omega_1$, respectively.

Figure 2 depicts the absorption signals of the EPR Rabi oscillations of the $E_1'$ centers detected for a fixed MW field amplitude ($\omega_1/2\pi = 1.030$ MHz) without the RF field and with the RF field. The observed signals demonstrate the frequency shift of the Rabi oscillations in the bichromatic field. The shift agree well with our theoretical predictions provided by equation (12) for $T_2 = 3.5$ μs. Because $T_2 \ll T_1 = 0.2$ ms, the influence of the longitudinal relaxation is neglected. In accordance with equation (12), the frequency shift of the Rabi oscillations does not change their decay rate. Thus, only the relaxation times limit the observed effect of the higher-order geometric phase.

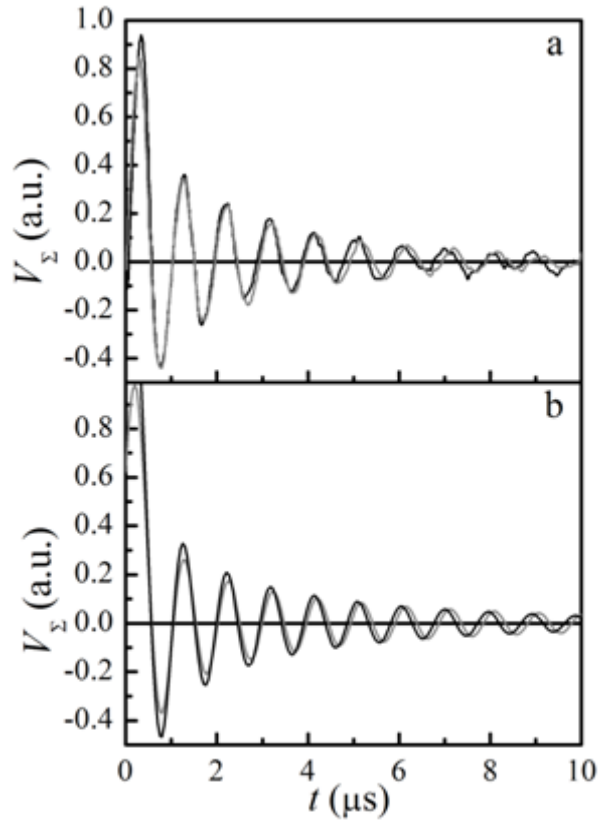

Figure 2. The Rabi oscillations of the $E_1'$ centers in crystalline quartz without and with the RF field. a) The signals detected at $\omega_{mw} = \omega_0$ and $\omega_1/2\pi = 1.030$ MHz: for the gray line $\omega_2 = 0$, for the black line $\omega_2/2\pi = 0.123$ MHz, $\omega_{rf}/2\pi = 0.500$ MHz b) Results of analytical calculations based upon equation (12) for the parameters used in the experiment, with color coding of lines as in a). The frequency shift of the Rabi oscillations is 0.017 MHz.

The frequency shift of the Rabi oscillations as a function of the RF frequency for the different values of the RF field amplitude at the fixed MW field amplitude ($\omega_1/2\pi = 1.030$ MHz) is displayed in figure 3. The Rabi frequency shift was obtained by fitting the experimental data with equation (12). The dashed, dotted and solid lines in the figure are the second-order, fourth-order and exact calculations for the parameters of the bichromatic field used in our experiment. The experimental results are in very good agreement with the theoretical predictions.

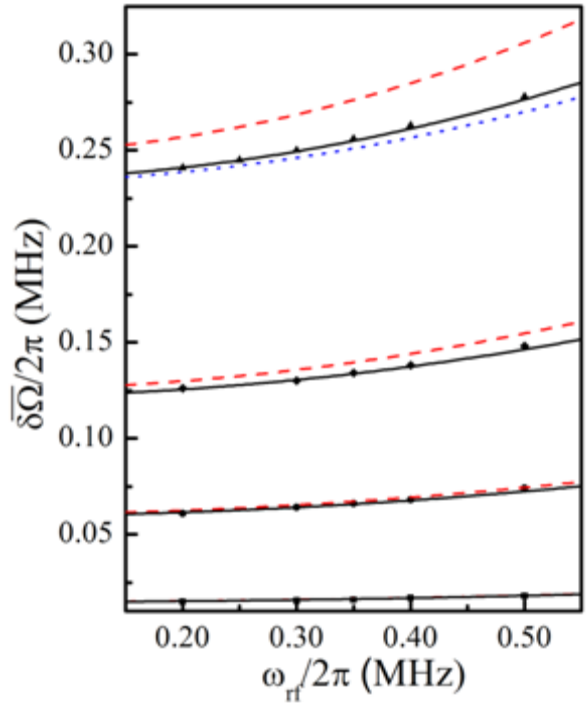

Figure 3. The frequency shift of the Rabi oscillations vs the frequency of the RF field at different RF amplitudes: = 0.123 MHz (squares), 0.249 MHz (circles), 0.359 MHz (diamonds), 0.505 MHz (triangles). and = 1.030 MHz. The dashed, dotted and solid lines are the second-order, fourth-order and exact calculations.

Figure 4 depicts the dependence of the measured geometric phase on the solid angle for the data presented in figure 3. The solid angle was varied by changing of the frequency and amplitude of the RF field. The experimental values of the geometric phase were obtained from equation (10) in the form $\varphi_g = (2\pi/\omega_{rf})(\overline{\delta\Omega}_{exp} - 2[(\omega_1^2 + \omega_2^2)^{1/2} - \omega_1])$, where $\overline{\delta\Omega}_{exp}$ is the measured shift of the Rabi frequency. The obtained results are in a good agreement with the predicted dependence given by $\varphi_g = \Theta'_C/2$ and presented by the solid line.

In the recent experiment with solid-state qubits [17] it was demonstrated that the non-adiabatic higher-order geometric phase originating from low-frequency fluctuations in the qubit transition frequency results in the geometric dephasing at the measurement of the Berry phase. In our experiment only the higher-order geometric phase is observed. Moreover, the observable dephasing is due to the conventional relaxation time $T_2$, and the geometric dephasing is eliminated by the stability of the bichromatic field.

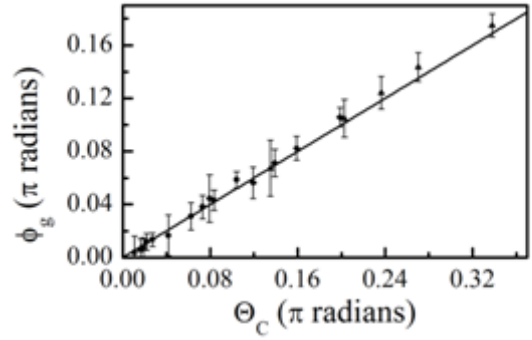

Figure 4. The higher-order geometric phase vs the solid angle at different RF amplitudes: = 0.123 MHz (squares), 0.249 MHz (circles), 0.359 MHz (diamonds), 0.505 MHz (triangles). = 1.030 MHz. The solid line is the predicted dependence given by .

**4. Summary**

We have shown both theoretically and experimentally that, in the evolution of a spin qubit driven by bichromatic (microwave and radiofrequency) field, the geometric phase during the full period of the "slow" RF field appears as a shift of the Rabi frequency of the qubit in the MW field. We have demonstrated that for the linearly polarized RF field, only the higher-order geometric phase is observed. This phase can be realized in a variety of quantum systems, thus opening new possibilities for their geometric control, and our results can be useful in any practical applications beyond ideal adiabaticity.